\documentstyle[aps,twocolumn]{revtex}

\begin{document}

\bibliographystyle{unsrt}
\title{Hamiltonian formalism and the
Garrett-Munk spectrum of internal waves in the ocean
}
\author {Yuri V. Lvov\cite{yuri} \and  Esteban G.Tabak
\cite{newell}} \address{ \cite{yuri}~~ 
Department of Mathematical Sciences, Rensselaer Polytechnic Institute,
Troy, NY 12180 \\
 \cite{newell}~~
Courant Institute of Mathematical Sciences,
        New York University, New York, NY 10012. \\
{\bf This paper first appeared at PRL Vol 87, page 168501, 2001}\\
{\it and contained some misprints corrected below}
} \maketitle
\maketitle


\begin{abstract} 
Wave turbulence formalism for long internal waves in a stratified
fluid is developed, based on a natural Hamiltonian description. A
kinetic equation appropriate for the description of spectral energy
transfer is derived, and its self-similar stationary solution
corresponding to a direct cascade of energy toward the short scales is
found. This solution is very close to the high wavenumber limit of the
Garrett-Munk spectrum of long internal waves in the ocean. In fact, a
small modification of the Garrett-Munk formalism includes a spectrum
consistent with the one predicted by wave turbulence.\end{abstract}


{\it {Introduction.}---} Remarkably, the internal wave spectrum in the
deep ocean has much the same shape wherever it is observed, except
when the observations are made close to a strong source of internal
waves \cite{Gill}. This observation led Garrett and Munk
\cite{GM72,GM75,GM79} to propose an analytical form of the internal
wave spectrum that approximates many observations. This spectrum is
now called the Garrett-Munk (GM) spectra of internal waves. The total
energy of the internal waves may be represented as an integral over
spectral energy density\footnote{A number of missprints in the
original PRL {\bf 87} 168501 (2001) manuscript stemed from the fact
that GM spectra is a one dim spectra, i.e. to obtain total energy one
should integrate it over $k$ and $m$, not $\bf k$ and $m$ as we
mi
stakenly thought - Thank you to Dr. Kurt Polzin for noticing this to
us.}
\begin{eqnarray}
E=\int E({\bf k},m) \, d { k} \, d m  \, ,
\end{eqnarray}
where ${k}$ and $m$ are the horizontal and vertical components of the 
wavevector respectively. 
Under the assumption of horizontal isotropy, Garrett and
Munk proposed the following empirical expression for the spectral
energy density:
\begin{eqnarray}\label{GM}
  E(k,m)= \frac{2\, f \, N \, E  \,
     m/m^*}{\pi \left(1+\frac{m}{m^*}\right)^{5/2} \left(N^2 k^2 + f^2
     m^2\right)} \, . 
\end{eqnarray}
Here $E$ is a constant, quantifying the total energy content of the
internal wave spectrum, $f$ is the Coriolis parameter, $N$ is the buoyancy
frequency, $k=|{\bf k}|$,
and $m^*$ is a reference vertical wavenumber to be determined
from observations.

The dispersion relation underlying this proposed spectrum is that
of long internal waves, with a profile rapidly oscillating in the vertical, 
so that both the hydrostatic balance and the WKB approximation apply:
\begin{equation}
  \omega^2 = f^2 + N^2 \left({k}/{m} \right)^2 \, .
  \label{DR}
\end{equation}
Using this dispersion relation, the spectrum can be
transformed from wave-number space $(k,m)$ into 
frequency-horizontal wavenumber space $(k,\omega)$, 
or frequency-vertical wave
number space $(m,\omega)$. In particular, the integral
of $E(k,\omega)$ over $k$ --or equivalently, the integral of
$E(m,\omega)$ over $m$-- yields the moored spectrum
\begin{equation}
  E(\omega) = 
   {2\, f\, E}\left(
   {\pi \left(1-({f}/{\omega})^2\right)^{1/2}\omega^2}
\right)^{-1}
 \, ,
   \label{moored}
\end{equation}
with an $1/\omega^2$ dependence away from the inertial frequency 
that appears prominently in moored observations.

The GM spectrum constitutes an invaluable tool for oceanographers,
assimilating hundreds of different observations into
a single, simple formula, that clarifies the distribution of the
energy contents of internal waves among spatial and temporal scales.
Yet a number of questions regarding the spectrum itself remain open.
One is about its accuracy: Garrett and Munk basically made up a
family of spectra, depending on a few parameters, that was simple
and consistent with the dispersion relation (\ref{DR}) for
long internal waves; and then fitted the parameters to match
observational data. Their success speaks of their 
powerful intuition, yet it leaves the door open
to question both the accuracy of the parameter fit, and the
appropriateness of the proposed family of spectra, necessarily
incomplete, which involved
a high degree of arbitrariness. The other question, more fundamental
to theorists, is to explain or derive the form of the spectrum from
first principles. Such explanation should surely involve the nonlinear
interaction among internal modes, as well as the nature of the forcing
and dissipation acting on the system.

In this work, we elucidate which spectrum the theory of wave
turbulence (WT) would predict for internal waves in scales far away from both
the forcing and the dissipation. Wave turbulence theory (also called
weak turbulence, to contrast it to the ``strong'' turbulence of
isotropic fluids) applies to Hamiltonian systems characterized by
a scale separation between a fast, linear dispersive wave structure, 
and its slow, nonlinear modulation.

In what follows we assume that there is pumping of energy into the
internal wave field by the wind, by interaction with surface waves,
or by other processes.  We assume that these pumping processes can be
characterized by wavelengths of the order of hundreds of
meters. Moreover, we assume that the processes which
remove energy from the internal wave field, such as wave breaking,
turbulent mixing, multiple reflections
from the surface and bottom boundary layers, or interaction with bottom
topography, can be characterized by lengthscales of less than a
meter. Then there is a region of lengthscales, called {\it inertial
interval} or {\it transparency region}, where dissipation and pumping
processes are not important.  It is the nonlinear wave
interaction which determines the form of the spectrum in the region of
transparency. Energy then gets into the system of internal
waves at large scales, cascades through the inertial scales via
multiple nonlinear interactions,
and is absorbed in the small scales by the dissipation
processes. This scenario corresponds to the {\it Kolmogorov spectrum}
of WT theory~\cite{ZLF}.
According to WT, the energy distribution in
the inertial region is defined solely by the nature of nonlinear
interactions and by the linear dispersion relation of
the waves in the system.

Notice that, in the transparency region, the dispersion relation
is self-similar, since $|\omega| >> f$. When this is the case,
and the nonlinearity is homogeneous, WT theory predicts
self-similar stationary solutions.
For $|m| >> m^*$ and $|\omega| >> f$, the Garrett-Munk spectrum 
(\ref{GM}) becomes
\begin{equation}
E(k, m) \simeq \left(k^2 m^{3/2}\right)^{-1} \, . \label{GMLW}
\end{equation}
The spectrum we shall obtain below using the WT 
formalism is, instead,
\begin{equation}
E(k, m) \simeq \left( k^{3/2} m^{3/2} \right)^{-1} \, . \label{GMWT}
\end{equation}
The small disparity in the exponents of the Garrett-Munk spectrum
and the prediction of WT theory may be attributed
either to effects that WT does not capture, such as 
wave interaction with shear and vorticity and wave breaking,
or to inaccuracies of the GM spectrum itself. In the last
section of this letter, we introduce a slight modification of
GM, which yields the spectrum in (\ref{GMWT}). 

{\it{Hamiltonian structure and kinetic equation}}---
The equations for long internal waves in an incompressible
stratified fluid are
\begin{eqnarray}
\frac{d {\bf u}}{d t} + \frac{{\bf \nabla} P}{\rho} = 0,& \ \ \ & 
P_z+\rho g = 0, \nonumber\\
\frac{ d \rho}{d t} =0,& \ \ \ & 
{\bf \nabla}\cdot {\bf u} + w_z = 0 \, , \nonumber  
\end{eqnarray}
where ${\bf u}$ and $w$ are the horizontal and vertical components
of the velocity respectively, $P$ is the pressure, $\rho$ the density, 
$g$ the gravity constant, 
$\nabla = (\partial_x, \partial_y)$ the horizontal gradient
operator, and
$$
  \frac{d}{dt} = \frac{\partial}{\partial t} + 
    {\bf u}\cdot{\bf \nabla} + w \frac{\partial}{\partial z}
$$
is the Lagrangian derivative following a particle.
Notice that we are considering waves long  enough for the
hydrostatic balance to be valid, yet not so long to feel the
effects of the rotation of the earth. This is consistent with
the scales of the conjectured transparency region described above.

Changing to isopicnal coordinates, where the roles of the vertical
coordinate $z$ and the density $\rho$ as independent and dependent
variables are reversed, the equations become:
$$
\frac{\partial {\bf u}}{\partial t} +{\bf u} \cdot \nabla {\bf u}
+ \frac{\nabla M}{\rho} = 0, \ \ 
  M_{\rho} =g  z,\ \  
  z_{\rho t} + \nabla \cdot \left(z_{\rho} {\bf u} \right)
	 = 0 \, .  \nonumber 
$$
Here $\nabla = (\partial_x, \partial_y)$ is the isopicnal gradient,
acting along surfaces of constant density, and $M$ is the Montgomery
potential 
$  M=P+\rho\, z.  $
We shall consider flows which are irrotational along isopicnals;
for these, it is convenient to introduce a horizontal velocity potential
$\phi$, such that 
$ {\bf u}={\bf \nabla} \phi({\bf x}, \rho, t) \, , $
reducing the equations further to the pair
\begin{eqnarray}
  \phi_t + \frac{1}{2} |\nabla \phi|^2 + \frac{g}{\rho}
     \int^{\rho}\int^{\rho_2} \frac{\Pi-\Pi_0}{\rho_1} \, d\rho_1 \, d\rho_2
     &=& 0 \, , \label{H1}
	 \\ \Pi_t + \nabla \cdot \left(\Pi\, \nabla \phi \right)
     &=& 0 \, . \label{H2}
\end{eqnarray}
Here we have introduced the variable $\Pi = \rho M_{\rho \rho}/g = \rho \,
z_{\rho} $, and, for future convenience, a reference equilibrium
value $\Pi_0(\rho) = -g/N^2$, where $N^2=-g \, \rho_0'(z)/\rho_0$ 
is the square of the buoyancy frequency. 
The variable $\Pi$, representing the stratification
lengthscale, is the canonical conjugate of $\phi$ under the
Hamiltonian flow given by
\begin{equation}
  {\cal H} =  \frac{1}{2}\int \left( \Pi \, |\nabla \phi|^2 -
   g \left|\int^{\rho} \frac{\Pi-\Pi_0}{\rho_1} \, d\rho_1\right|^2 \right)\,
    d{\bf x} \, d\rho \, .
\label{TrueHam}
\end{equation}
The first term in this Hamiltonian clearly corresponds to the
kinetic energy of the flow; that the second term is in fact the
potential energy follows from the simple calculation
\begin{eqnarray}
\frac{1}{2} \left|\int^{\rho} \frac{\Pi-\Pi_0}{\rho_1} \, d\rho_1\right|^2
d\rho =
\frac{1}{2}  \left|\int^{z} (dz-dz_0)\right|^2 d\rho 
= \nonumber\\ \frac{1}{2} (z-z_0)^2 d\rho = - \rho \, (z-z_0) \, dz +
d \left(\frac{1}{2} \rho\, (z-z_0)^2 \right) \, ,
\end{eqnarray}
so that the second term in (\ref{TrueHam}) is simply $
g\int \rho \, (z-z_0) \, dz  \, .$
The equations of motion (\ref{H1}, \ref{H2}) can be written
in terms of the Hamiltonian (\ref{TrueHam}) in the canonical form
\begin{equation}
\Pi_t=\frac{\delta {\cal H}}{\delta \phi}, 
\ \ \ \  \ \ \ 
\phi_t=-\frac{\delta {\cal H}}{\delta \Pi} \, .
\label{Canonical}
\end{equation}

For simplicity, we shall take the buoyancy frequency $N$ of the
equilibrium profile to be a constant, and we shall replace the
density $\rho$ in the denominator of the Hamiltonian's potential
energy by a constant $\rho_0$. This is the WKB --also Boussinesq-- 
approximation, which makes sense for waves varying rapidly in the
vertical direction, particularly since the water density 
typically changes only by a few percent over the full depth of
the ocean.

Let us decompose $\Pi$ into the
sum of its equilibrium value and deviation from equilibrium 
$\Pi=\Pi_0+\Pi' \, .$
Since, in the WKB limit, the linear part of the resulting Hamiltonian 
has constant coefficients, it is natural to perform a Fourier
transformation in both vertical and horizontal directions. We assume
that both Fourier spectra are continuous, which is a reasonable
approximation if the wave's vertical wavelengths are much smaller 
than the depth of the ocean. 
Then
\begin{eqnarray}
{\cal H}=\int \left(-\frac{g k^2}{2 N^2}|\phi_{\bf p}|^2 -
\frac{g}{2 \rho_0^2 m^2} | \Pi_{\bf p}'|^2\right)\, d {\bf p} \nonumber
-\\ \frac{1}{(2\pi)^{3/2}}
\int {\bf k_2}\cdot {\bf k_3} \, \Pi_{\bf p_1}' \phi_{\bf p_2}
\phi_{\bf p_3}^* \, \delta( {\bf p_1} + {\bf p_2}
- {\bf p_3} ) \, d {\bf p}_{123} \, , \nonumber\\
\label{Hamiltonian}
\end{eqnarray} where ${\bf p}$ is a
three-dimensional wave vector ${\bf p}=\{{\bf k},m \}$ and
$ d {\bf p}_{123}\equiv d {\bf p}_{1} d {\bf p}_{2} d {\bf p}_{3}$. 
 
We now introduce a canonical transformation which transfers the
equations of motion for the Fourier components $ \phi_{\bf
p}$ and $\Pi_{\bf p}'$ into a single equation for the canonical
variable $a_{\bf p}$. We choose this transformation
so that the quadratic part of Hamiltonian becomes diagonal in
$a_{\bf p}$. The canonical transformation reads:
\begin{eqnarray}
\phi_{\bf p}=i\sqrt{\frac{N}{2 \rho_0 m k}}(a_{\bf p}-a_{\bf -p}^*
), \ \ \  \Pi'_{\bf p} = \sqrt{\frac{\rho_0 m k}{2 N}}(a_{\bf
p}+a_{\bf -p}^* ), \ \ \ \nonumber
\end{eqnarray} 
Then the pair of the canonical equations of motion (\ref{Canonical})
become the single equation
\begin{equation}
i\frac{\partial}{\partial t} a_{\bf p} = 
  \frac{\partial {\cal H}}{\partial a_{\bf p}^*} \, ,
\end{equation}
with Hamiltonian
\begin{eqnarray}
{\cal H}=\int \omega_p \, |a_{\bf p}|^2 \, d {\bf p} + \nonumber \\
\int 
 V_{ {\bf p_1} {\bf p_2} {\bf p_3}} \left( a_{\bf p_1}^* a_{\bf
p_2}^* a_{\bf p_3} + a_{\bf p_1} a_{\bf p_2} a_{\bf p_3}^*\right)\,
\delta_{ {\bf p_1} + {\bf p_2} - {\bf p_3}} \, d{\bf p}_{123} +\nonumber \\
\int 
 V_{ {\bf p_1} {\bf p_2} {\bf p_3}} \left( a_{\bf p_1}^* a_{\bf
p_2}^* a_{\bf p_3}^* + a_{\bf p_1} a_{\bf p_2} a_{\bf p_3}\right) \,
\delta_{ {\bf p_1} + {\bf p_2} +{\bf p_3}}\, d {\bf p}_{123}  .\nonumber\\
\label{HAM}
\end{eqnarray}
Here 
$\omega_{\bf p}$ is the linear dispersion 
relation for the Hamiltonian (\ref{Hamiltonian}), 
$$ \omega_{\bf p}\equiv \omega_{{\bf k},m} = 
   \frac{g}{N \rho_0} \frac{k}{|m|} \, ,$$
and $ V_{ {\bf p_1} {\bf p_2} {\bf p_3}}$ is the internal wave
interaction matrix element, given by
\begin{eqnarray}
V_{ {\bf p_1} {\bf p_2} {\bf p_3}}=\sqrt{ |{\bf k_1}| |{\bf k_2}|
|{\bf k_3}|} \, 
\left( \frac{ {\bf k_1}\cdot{\bf k_2}}{
|{\bf k_1}| |{\bf k_2}|} \sqrt{\left|\frac{m_3}{m_1 m_2}\right|} 
\right.
\nonumber \\
\left.
+ \frac{ {\bf
k_1}\cdot{\bf k_3}}{ |{\bf k_1}| |{\bf k_3}|} \sqrt{\left|\frac{m_2}{m_1
m_3}\right|} + \frac{ {\bf k_2}\cdot{\bf k_3}}{ |{\bf k_2}| |{\bf k_3}|}
\sqrt{\left|\frac{m_1}{m_2 m_3}\right|} \right)\, . 
\nonumber
\end{eqnarray}

The form of this Hamiltonian is typical for systems with 
three-wave interactions and cylindrical symmetry. Following
wave turbulence theory, one proposes a perturbation expansion
in the amplitude of the nonlinearity.
To zeroth order in the perturbation, one recovers the linear waves. 
At higher orders, the nonlinear interactions lead to a slow modulation
of the wave amplitudes, representing spectral
transfer of the conserved quantities of the Hamiltonian. 
This transfer manifests itself in the perturbation expansion through {\it
resonances} or {\it secular terms}, occurring on the
so called {\it resonant manifold}.  Energy transfer is described by an
approximate {\it kinetic equation} for the ``number of waves'' or wave-action
$n_{\bf p}$, defined by 
$$n_{\bf p} \delta({\bf p} - {\bf p'}) = \langle a_{\bf p}^* a_{\bf
p'}\rangle \, .$$ 
This kinetic equation is the classical analog of the Boltzmann
collision integral; it has been used for describibg surface water
waves since pioneering works by Hasselmann~\cite{Hass} and
Zakharov~\cite{Z68a,Z68b}. The derivation of the kinetic equation using the
wave turbulence formalism can be found, for instance, in \cite{ZLF}.
For the three-wave Hamiltonian (\ref{HAM}), the kinetic equation reads:
\begin{eqnarray}
\frac{d n_{\bf p}}{dt} = \pi \int 
 |V_{p p_1 p_2}|^2 \, f_{p12} \,
\delta_{{{\bf p} - \bf{p_1}-\bf{p_2}}} \, \delta_{\omega_{{\bf p}} 
-\omega_{{\bf{p_1}}}-\omega_{{\bf{p_2}}}} 
d {\bf p}_{12} \, , 
\nonumber \\ 
-2\pi\int
 \, |V_{p_1 p p_2}|^2\, f_{1p2}\, \delta_{{{\bf p_1} - \bf{p}-\bf{p_2}}} \,
  \delta_{{\omega_{{\bf p_1}} -\omega_{{\bf{p}}}-\omega_{{\bf{p_2}}}}}\Big)
\, d {\bf p}_{12} \, , 
\label{KEinternal}
\end{eqnarray}
where $ f_{p12} = n_{{\bf p_1}}n_{{\bf p_2}} - 
n_{{\bf p}}(n_{{\bf p_1}}+n_{{\bf p_2}}) \, .
$

Assuming horizontal isotropy, one can average
(\ref{KEinternal}) over all horizontal angles, obtaining
\begin{eqnarray} \label{KEinternalAveragedAngles}
 \frac{d n_p}{d t} 
= \frac{1}{k}\int 
\left(R^k_{12} - R^1_{k2} - R^2_{1k} \right) \, 
d k_1 d k_2 d m_1 d m_2 \, , \nonumber \\
R^k_{12}=\Delta^{-1}_{k 1 2} \, 
   \delta(\omega_{p}-\omega_{p_1}-\omega_{p_2}) \,
f^k_{12} \, |V^k_{12}|^2 \, \delta_{m-m_1-m_2} k k_1 k_2
\, , \nonumber \\
 \Delta^{-1}_{k 1 2} = \left< \delta({\bf k}-{\bf
k_1}-{\bf k_2})\right>\equiv \int \delta({\bf
k}-{\bf k_1}-{\bf k_2}) \, d \theta_1 d \theta_2 \, , \nonumber\\
\Delta _{k 1 2} = \frac{1}{2}\sqrt{
2 \left( (k k_1)^2 +(k k_2)^2 +(k_1 k_2)^2 
\right)-k^4-k_1^4 -k_2^4} \, .
\end{eqnarray}

In wave turbulence theory, three-wave kinetic equations admit two classes 
of exact stationary solutions: thermodynamic equilibrium and 
Kolmogorov flux solutions, with the latter
corresponding to a direct cascade of energy
--or other conserved quantities-- toward the higher modes.
The fact that the thermodynamic equilibrium --or equipartition of energy--
$n_{\bf
p} = 1/\omega_{\bf p}$ is a stationary solution of
(\ref{KEinternalAveragedAngles}) can be seen by
inspection, whereas in order to find Kolmogorov spectra one needs to
be more elaborate.  Let us assume that $n_{\bf p}$ is given by
the power-law anisotropic distribution
\begin{equation}\label{n}  
  n_{{\bf k},m}= k^x |m|^y \, .
\end{equation}
We will find exponents $x$ and $y$ by requiring that (\ref{n}) is a 
stationary solution to (\ref{KEinternalAveragedAngles}).
We shall use a version of Zakharov's transformation
\cite{Z68a,Z68b} introduced for cylindrically symmetrical systems by
Kuznetsov in~\cite{Kuzia}.  Let us subject  the integration variables in the
second term $R^1_{k2}$ in (\ref{KEinternalAveragedAngles}) to the 
following transformation:
\begin{eqnarray}\nonumber
k_1 = {k^2}/{k_1'}, \  
m_1={m^2}/{m_1'}, \ 
k_2 = {k k_2'}/{k_1'}, \  m_2 = {m m_2'}/ {m_1'}.
\nonumber\end{eqnarray} Then $R^1_{k2}$ becomes $R^k_{12}$ multiplied
by a factor 
$${\left(\frac{k_1}{k}
\right)^{-6 - 2 x}
\left(\frac{m}{m_1} 
\right)^{2 + 2y}}.$$ Furthermore, let us subject
the third term $R^2_{1k}$ in (\ref{KEinternalAveragedAngles}) to the
following conformal transformation:
\begin{eqnarray} \nonumber
k_1 = {k k_1' }/{k_2'},
 \   m_1={m m_1'}/{m_2'},
\   
k_2 = {k^2}/{k_2'},  \ m_2 = {m^2}/{m_2'}.
\end{eqnarray}
Then  $R^2_{1k}$ becomes $R^k_{12}$ multiplied by a factor
$$\left(\frac{k_2}{k}\right)^{-6 - 2 x} 
\left(\frac{m}{m_2} \right)^{2 + 2y}.$$
Therefore   (\ref{KEinternalAveragedAngles}) can be written as
\begin{eqnarray} \label{KEtransformed}\frac{d n_{\bf p}}{d t} 
= \frac{1}{k}\int R^k_{12} \,
\left(1-
\left(\frac{k_1}{k}\right)^{-6 - 2 x}
 \left(\frac{m}{m_1} \right)^{2 + 2y}-\right. \nonumber \\ \left. 
\left(\frac{k_2}{k}\right)^{-6 - 2 x} 
\left(\frac{m}{m_2} \right)^{2 + 2y}
\right) \,  d k_1 d k_2 d m_1 d m_2 \, . \label{KEZtransformed}
\end{eqnarray}
We see that the choice $ -6- 2 x = 2 + 2 y = 1 \, ,$
which gives
$x=-{7}/{2}, \ \ \ \ y=-{1}/{2} \, ,$
makes the right-hand side of
(\ref{KEinternalAveragedAngles}) vanish due to the delta function 
in the frequencies, corresponding to energy conservation. 
The resulting wave action and spectral energy
distributions are given by
\begin{eqnarray} n_{{\bf k},m}= |{\bf k}|^{-7/2}
 |m|^{-1/2}, \ \ \\
E_{{k},m} =  k \omega_{{\bf k},m} n_{{\bf k},m}= |{\bf
k}|^{-3/2} |m|^{-3/2},\nonumber\end{eqnarray} 
This solution, ((\ref{GMWT}) of the introduction) corresponds
to the flux of energy from the large to the small scales.

This is the main result of this article - the short wave part of the
Garrett-Munk spectra is close to the stationary solution to
a kinetic equation hereby derived for internal
waves, based on a Hamiltonian structure appearing naturally
in isopicnal coordinates.

{\it{A modified Garrett-Munk spectrum}}---
In this section, we show how a minor modification of the
Munk-Garrett formalism allows us to match the wave turbulence
prediction (\ref{GMWT}). As in \cite{GM75}, we introduce two functions
$$ A(\lambda) = \frac{t-1}{(1+\lambda)^t} , \ \ \ 
 B(\omega) = \frac{2 f}{\pi}
\frac{1}{\left(1-\left(\frac{f}{\omega}\right)^2\right)^{1/2}
   \, \omega^2} \, ,
$$
and a reference, frequency dependent wave vector $(k^*,m^*)$.
In \cite{GM75}, $m^*$ was a constant, and $k^*$ was given
by $k^*=(w^2-f^2)^{1/2} m^*$. 
However, the same formalism carries
through if one allows both $k^*$ and $m^*$ to depend on $\omega$,
provided that the condition
$ {k^*}/{m^*} = \sqrt{w^2-f^2} $
is met. In keeping with the spirit of self-similarity, we shall
propose that
\begin{equation}
  k^* = \gamma \left(\omega^2-f^2\right)^{\frac{1-\delta}{2}}
  \, , \qquad
  m^* = \gamma \left(\omega^2-f^2\right)^{-\frac{\delta}{2}} \, ,
\end{equation}
where $\delta$, like $t$ and $\gamma$, is a constant to be determined from 
observations (or, in our case, from wave turbulence theory).

Then, following \cite{GM75}, we propose an energy spectrum
of the form
$$ E(k,\omega) = E \, A\left(\frac{k}{k^*}\right) \frac{B(\omega)}{k^*} \,
.$$
It follows from the dispersion relation that we have
\begin{eqnarray}
E(m,\omega) = E \, A\left(\frac{m}{m^*}\right) \frac{B(\omega)}{m^*} \, \\
 E(k,m) = \frac{2 \, f \, N \, E}{\pi}
    \frac{(m/m^*) \, A(m/m^*)}{N^2 k^2 + f^2 m^2} \, .
\label{GMgeneralized}
\end{eqnarray}
If we pick $t=2$ and $\delta=-1/2$, the asymptotic behavior of 
(\ref{GMgeneralized}) agrees with the prediction (\ref{GMWT})
of wave turbulence theory. In \cite{GM75}, $\delta$ was zero
by default, and $t=2.5$. Notice that, independently of the choices
of $t$ and $\delta$, the moored spectrum is always given by
(\ref{moored}).

{\it{Conclusion.}}--- We have found a natural Hamiltonian formulation
for long internal waves, and used it within the wave turbulence
formalism to determine the stationary energy spectrum corresponding to
a direct cascade of energy from the long to the short waves. This
spectrum is close to the one that Garrett and Munk fitted to available
observational data.  The small difference could be due either to
physical effects that the wave turbulence formalism fails to capture,
or to a real necessary correction to the GM spectrum.  We show how a
slight modification of the GM spectrum yields results in agreement
with WT theory.


\begin{thebibliography}{1}
\bibitem[\ast]{yuri} Yu.V. Lvov's e-mail: lvovy@rpi.edu
\bibitem[\ddag]{newell}E. Tabak's e-mail: tabak@cims.nyu.edu
\bibitem{Gill}
Gill.
 {\em Atmosphere-Ocean Dynamics}.
 Academic Press, 1982.
\bibitem{GM72}
Garrett C.J.R and Munk W.H.
 Space time scales of internal waves.
 {\em Geophysical Fluid Dynamics}, 3:225--264, 1972.
\bibitem{GM75}
Garrett C.J.R and Munk W.H.
 Space time scales of internal waves, progress report.
 80:281--297, 1975.
\bibitem{GM79}
Garrett C.J.R and Munk W.H.
 Internal waves in the ocean.
 {\em Annual Review of Fluid mechanics}, 11:339--369, 1979.
\bibitem{ZLF}
V.E. Zakharov, V.S. L'vov, and G.Falkovich.
 {\em Kolmogorov Spectra of Turbulence}.
 Springer-Verlag, 1992.
\bibitem{Hass}K.Hasselmann ,''On the nonlinear energy transfer in a
gravity wave spectrum.''  Part $I$. General theory {\em J. Fluid
Mech.},12:481, 1962; \\"Part $II$. 
Conservation theorems, wave-particle analogy, irreversibility.'',
15:324, 1962.
\bibitem{Z68a}
V.E.Zakharov.
 Stability of periodic waves of finite amplitude on a surface of deep
  fluid.
 {\em J.Appl.Mech. Tech. Phys}, 2:190--198, 1967.
\bibitem{Z68b}
V.~E. Zakharov.
 The instability of waves in nonlinear dispersive media.
 {\em Sov. Phys. JETP}, 24(4):740--744, 1967.
\bibitem{Kuzia} E.A.Kuznetsov. {\em Xh. Eksp.Teor.Fiz}, 62,584, 1972.
\end{thebibliography}

\end{document}